\documentclass[letter,twocolumn]{jpsj2} 

\title{Pseudogap Formation and Heavy Carrier Dynamics in 
Intermediate Valence YbAl$_3$} 

\author{%
Hidekazu \textsc{Okamura}\thanks{E-mail: okamura@phys.sci.kobe-u.ac.jp}, 
Takahiro \textsc{Michizawa}, 
Takao \textsc{Nanba} and 
Takao \textsc{Ebihara}$^1$
}

\inst{%
Graduate School of Science and Technology, Kobe University, 
Kobe 657-8501. 
\\
$^1$Department of Physics, Shizuoka University, 
Shizuoka 422-8529.
}

\recdate{\today}

\abst{%
Infrared optical conductivity [$\sigma(\omega)$] of the 
intermediate valence compound YbAl$_3$ has been measured 
at temperatures 8~K $\leq T \leq$ 690~K to study its 
microscopic electronic structures.    
Despite the highly metallic characters of YbAl$_3$, 
$\sigma(\omega)$ exhibits a clear pseudogap (strong 
depletion of spectral weight) of about 60~meV below 40~K.  
It also shows a strong mid-infrared peak centered 
at $\sim$ 0.25~eV.    
Energy-dependent effective mass and scattering rate of 
the carriers obtained from the data indicate the 
formation of a heavy-mass Fermi liquid state.  
These characteristic results are discussed in terms of 
the hybridization states between the Yb 4$f$ and the 
conduction electrons.     
It is argued, in particular, that the pseudogap and the 
mid-infrared peak result from the indirect and the direct 
gaps, respectively, within the hybridization state.  
}

\kword{
YbAl$_3$, strongly correlated electron system, 
intermediate valence, heavy fermion, optical conductivity
}

\begin{document}
\maketitle

The duality and crossover between localized and itinerant 
characters exhibited by the 4$f$ electrons in rare-earth 
compounds, most typically those containing a lattice of 
Ce or Yb ions, has been a central issue in the physics of 
strongly correlated electron systems.\cite{sces}      
Such duality results from a hybridization of conduction ($c$) 
electrons having spatially extended wave functions and $f$ 
electrons having localized orbitals.   At 
temperatures $T \simeq T_K$, where $T_K$ is the single-site 
Kondo temperature, the $f$ electrons are localized well at 
each site.  The Kondo effect at individual sites give rise 
to a $\rho_m \propto -\log T$ behavior, where $\rho_m(T)$ 
is the magnetic contribution to the resistivity.      
In contrast, below another characteristic temperature $T^\ast$, 
where $T^\ast \ll T_K$ in general, the $f$ electrons become 
partly itinerant as a ``heavy quasiparticle band'' is formed.    
This is a spatially extended state consisting of both $c$ and 
$f$ electron wave functions with large effective masses 
($m^\ast$), which shows Fermi liquid properties such as 
$\rho_m \propto T^2$ dependence below $T^\ast$.     
($T^\ast$ is sometimes referred to as the coherence 
temperature.)    The magnetic susceptibility, $\chi(T)$, shows 
a maximum at $T = T_{max}$, marking a crossover between the 
local moment (Curie Weiss) regime above $T_{max}$ and the 
free carrier (Pauli paramagnetism) regime below $T_{max}$.     
The values of $m^\ast$ and $T_{max}$ are measures of the 
hybridization strength.    
``Heavy fermion'' (HF) compounds have only a weak hybridization, 
and show large $m^\ast$ values (up to $\sim$ 1000~$m_0$) 
and $T_{max}$ of typically a few K or lower.     
In contrast, ``intermediate valence'' (IV) compounds have 
a stronger hybridization, showing typically $m^\ast$ = 10-100~$m_0$, 
$T_{max}$ = 20-300~K, and the Ce or Yb valence away from 3+.   
In addition to the metallic cases described above, there are a 
small number of ``Kondo semiconductors'', where the $c$-$f$ 
hybridization leads to semiconducting 
characteristics.\cite{fisk,riseborough}

Although the above scenario is generally accepted, 
the microscopic mechanism for the $c$-$f$ hybridization 
and its consequences are not yet precisely known.   
Optical conductivity technique has emerged as a powerful 
tool for studying the microscopic electronic structures 
in $f$ electron systems.\cite{degiorgi}   It has been 
applied to many HF and IV compounds, including metals 
(CeCu$_6$ \cite{degiorgi}, CePd$_3$ \cite{sievers}), 
Kondo semiconductors (SmB$_6$\cite{wachter}, 
Ce$_3$Bi$_4$Pt$_3$\cite{schle}, 
YbB$_{12}$\cite{YbB$_{12}$-1,YbB$_{12}$-2}), and more recently 
filled-skutterudite compounds.\cite{basov,matunami}

In this work we have applied the optical technique to 
YbAl$_3$, which is a well-known IV compound extensively 
studied for the last few decades.   Recently, the first 
observation of the de Haas-van Alphen (dHvA) effect in 
YbAl$_3$ was reported,\cite{ebihara-Physica,ebihara-JPSJ} 
and YbAl$_3$ is gaining renewed 
interest~\cite{cornelius,ebihara-PRL,bauer}.    
The $m^\ast$ values obtained by dHvA are 14-24~$m_0$, the specific heat 
coefficient is $\gamma \sim$ 40~mJ/mol$\cdot$K$^2$, 
$T^\ast \sim$ 35~K and $T_{max} \sim$ 120~K.    
Experimentally estimated single-site $T_K$ and the Yb mean 
valence are in the ranges 600-700~K (50-60~meV) and 2.7-2.8, 
respectively.\cite{cornelius,PES}    
These results suggest a rather strong $c$-$f$ hybridization in 
YbAl$_3$.    The measured optical conductivity spectrum 
$\sigma(\omega)$ of YbAl$_3$ has shown strong $T$ 
dependences,\cite{krakov} which are analyzed based on 
the $c$-$f$ hybridization scenario within the framework 
of the periodic Anderson model.

YbAl$_3$ single crystals used in this work were grown with 
a self-flux method~\cite{ebihara-Physica}.    Samples taken 
from the same batch showed clear dHvA oscillations, 
indicating a high quality.    $\sigma(\omega)$ spectra were 
obtained from the measured optical reflectivity spectra 
$R(\omega)$ using the Kramers-Kronig relations.\cite{wooten}      
$R(\omega)$ was measured in the range 7~meV - 35~eV on 
mechanically polished surfaces of the samples, under a 
near-normal incidence using a thermal source and a synchrotron 
radiation source at BL7B of the UVSOR, Institute for 
Molecular Science.    
A gold or a silver film deposited {\it in situ} onto the 
sample surface  was used as a reference of the 
reflectivity.\cite{timusk}    To complete $R(\omega)$, 
a Hagen-Rubens formula was used for low-energy 
extrapolation,\cite{wooten} and an $\omega^{-4}$ function 
for high-energy extrapolation.

Figure 1(a) shows $R(\omega)$ of YbAl$_3$ at 295~K in the 
entire measured energy range, and Figure~1(b) shows the 
low-energy part of $R(\omega)$ at 8~K $\leq T \leq$ 690~K 
and that of non-magnetic LuAl$_3$ at 295~K.        
YbAl$_3$ has a broad dip in $R(\omega)$ below 0.5~eV, 
which becomes 
more pronounced with decreasing $T$.     In contrast, 
LuAl$_3$ has no such feature in $R(\omega)$.    
These results indicate that the dip in $R(\omega)$ of 
YbAl$_3$ is caused by the Yb 4$f$-related 
electronic states located near the Fermi level ($E_F)$.

Figure 2(a) shows $\sigma(\omega)$ of YbAl$_3$ over a 
wide energy range for 8~K $\leq T \leq$ 690~K, 
and Fig.~2(b) shows those in the infrared 
region below 295~K.     
$\sigma(\omega)$ spectra are characterized by the following 
three key features, all of which are strongly $T$-dependent: 
First, a steep rise of $\sigma(\omega)$ toward $\omega$=0 
is observed.   This is due to the Drude response of free 
carriers.\cite{wooten}      
Second, a pronounced mid-infrared (mIR) peak centered 
near 0.25~eV is observed, which becomes gradually stronger 
with decreasing $T$.   This peak results from the dip 
in $R(\omega)$ discussed above, and shows the existence 
of strong excitations due to the Yb 4$f$-derived 
electronic states near $E_F$.     
The third feature is the appearance below 120~K of a strong 
depletion (pseudogap) of $\sigma(\omega)$, indicated by the vertical 
arrow in Fig.~2(b), and the associated shoulder at 60~meV, 
indicated by the vertical broken line.    
The pseudogap formation in $\sigma(\omega)$ evidences that the 
density of states (DOS) near $E_F$ decreases below 120~K.   
Note that $T_{max}$ of YbAl$_3$ is also 120~K, 
i.e., the crossover from local moment to free carrier magnetism 
is coincident with the appearance of the pseudogap and shoulder 
in $\sigma(\omega)$.    
In addition, the shoulder position (60~meV) agrees well 
with $T_K$ of YbAl$_3$.  
$\sigma(\omega)$ shows large variation with 
$T$ above 40~K, but much less variation between 8 and 40~K.    
This is explicitly shown Fig.~2(c), which plots 
$\sigma(\omega)$ at the pseudogap 
and the integrated intensity of the mIR peak as a function of $T$.    
Namely, the evolution of the electronic structures with cooling 
is almost complete at 40~K.    This is consistent with the Fermi 
liquid behavior ($\rho \propto T^2$) of YbAl$_3$ observed below 
$T^\ast \simeq$ 35~K,\cite{ebihara-Physica,ebihara-JPSJ} i.e., 
the electronic structures do not change very much once a 
Fermi liquid state is established.  
The dc conductivity [$\sigma_{dc}$] and the low-energy 
region of $\sigma(\omega)$ are compared to each other in Fig.~2(d).   
With decreasing 
$T$, $\sigma(\omega)$ decreases while $\sigma_{dc}$ increases rapidly.   
From this plot, it is apparent that an extremely narrow 
Drude peak grows below the low-energy limit of our measurement, 
where at 8~K $\sigma(\omega)$ increases by more than two orders of 
magnitude within a range of several meV.

Figure~3 shows the energy-dependent effective mass 
$m^\ast(\omega)$ and the scattering rate 
$\tau^{-1}(\omega)$ of the carriers, obtained with the 
``generalized Drude'' 
analysis~\cite{basov2} to the present data.     
At 8~K, $\tau^{-1}(\omega)$ 
approaches zero with decreasing $\omega$ below $\sim$ 40~meV, 
with approximately following an $\omega^2$ dependence.    
An $\omega^2$ dependence of the scattering rate is a 
signature of the Fermi liquid.\cite{A-M}      
In contrast, $\tau^{-1}(\omega)$ does not approach zero 
as $\omega \rightarrow 0$ at higher $T$'s.       
This is consistent with the result that another Fermi liquid 
property of $\rho(T) \propto T^2$ is observed only below 
$\sim$ 35~K.\cite{ebihara-Physica}     
$m^\ast(\omega)$ below $\sim$ 40~meV becomes large with 
decreasing $T$, reaching about 30 times the bare band 
mass at 8~K.    This value is consistent with the 
cyclotron masses of 14-24~$m_0$ obtained by the dHvA 
experiments.\cite{ebihara-JPSJ}

Among the characteristic optical results of YbAl$_3$ presented 
above, the narrow Drude peak and the mIR peak are 
not peculiar to YbAl$_3$, but similar features have been 
widely observed for many HF/IV 
metals.\cite{degiorgi,degiorgi-mIR}  
Although less common, pseudogap formation 
and heavy-mass Fermi liquid properties have been also 
reported on some HF-IV compounds, including CePd$_3$.\cite{sievers}   
However, the present work is a rare case where all of these 
characteristic features are observed so clearly for a 
single compound.     
Note that the pseudogap in $\sigma(\omega)$ is not merely a tail of 
the mIR peak but is a distinct feature, since the former 
appears below 120~K ($\sim T_{max}$) and becomes 
well developed only at 40~K and below, while the 
latter is observed up to much higher temperatures.   
Namely, the evolution of $\sigma(\omega)$ appear to exhibit 
two different characteristic temperatures, as well as 
two different energy scales (the pseudogap below 60~meV 
and the mIR peak at 0.25~eV).     
The relation between these two features is never trivial.   
Below, we first describe how the narrow Drude peak and the 
mid-IR peak in HF/IV metals have been interpreted based on 
the periodic Anderson model (PAM).   Then we attempt to 
analyze the relation between the pseudogap and the mIR peak.

PAM has been a most common model to study theoretically 
the physical properties of HF/IV compounds.\cite{hewson,degiorgi} 
Since the PAM cannot be solved exactly, 
various approximate solutions of PAM have 
been reported,\cite{riseborough} including explicit 
calculations of the $T$-dependent $\sigma(\omega)$ based 
on the dynamical mean field 
theory.\cite{mutou,jarrel,rozenberg,logan}    
According to them, the electronic structures 
around $E_F$ at $T \ll T_K$ can be described by a pair of 
$c$-$f$ hybridized bands, with an associated energy gap, 
as illustrated in Fig.~4(a).     The flatness of the 
bands near $E_F$ (large $m^\ast$) results from the localized 
character of the (bare) $f$ electrons.    
For HF/IV metals, $E_F$ is located outside the gap.   
The resulting $f$-like, heavy quasiparticles 
can lead to a narrow Drude peak and Fermi liquid properties 
such as $\rho \propto T^2$ and $1/\tau \propto \omega^2$.\cite{millis}    
A semiconductor case is realized when $E_F$ is located 
within the gap,\cite{fisk,riseborough} but the gap 
formation mechanism is similar to that in the 
metallic case.     
In any case, the gap in the total DOS is an {\it indirect gap}, 
as sketched in Fig.~4(a), with the magnitude 
$\Delta_{ind} \sim k_B T_K$.       Optical transitions across 
$\Delta_{ind}$ are not allowed at low $T$'s due to the 
$k$ conservation rule.\cite{wooten}        
However, those across the {\it direct gap}, $\Delta_{dir}$, can 
occur as sketched in Fig.~4(a), resulting in a peak at 
$\omega=\Delta_{dir}$ in $\sigma(\omega)$, shown in Fig.~4(b).      
In fact the mIR peaks observed for HF/IV compounds have been 
commonly interpreted as arising from such 
transitions.\cite{degiorgi,degiorgi-mIR}    
The lack in the experimental $\sigma(\omega)$ of a step-like 
onset at $\Delta_{dir}$ sketched in Fig.~4(b) has been 
attributed to spectral broadening due to, e.g., impurities.

Clearly, the above model explains well the narrow Drude peak 
and the $\tau^{-1} \propto \omega^2$ dependence of YbAl$_3$ in 
this work.     Regarding the mIR peak and the pseudogap of 
YbAl$_3$, however, the above theory cannot be applied directly 
since it predicts only a single peak away from $\omega$=0 
[Fig.~4(b)].   
We attribute the mIR peak to optical transitions across 
$\Delta_{dir}$ and the pseudogap to {\it indirect transitions 
across $\Delta_{ind}$}.     
Although such indirect transitions are forbidden by the 
$k$-conservation rule as mentioned above, lattice disorder 
such as impurities may cause a partial relaxation of the rule, 
leading to indirect transitions.    
In addition, it has been pointed out\cite{riseborough,logan} 
that many-body scatterings, inherent in strongly-correlated 
systems, may provide the extra momentum needed for indirect 
transition.   This would cause a long tail of $\sigma(\omega)$ 
down to $\Delta_{ind}$ as sketched in Fig~4(c), which is very 
similar to that observed for YbAl$_3$.   
Our interpretation above is consistent with the following 
key experimental results and the theoretical predictions 
by the PAM:  
(i)The pseudogap width agrees well with $T_K$ of YbAl$_3$, 
$\sim$ 60~meV.\cite{cornelius,PES}   This is quite a strong 
evidence that the pseudogap is related with $\Delta_{ind}$.  
(ii)The mIR peak is centered at $\sim$ 0.25~eV, much larger 
than the pseudogap width.      
Theories\cite{mutou,jarrel,rozenberg,logan} show that 
$\Delta_{dir}$ can be several times larger than 
$\Delta_{ind}$, which is also seen from Fig.~4(a).    
The observed energies of the mIR peak and the pseudogap 
for YbAl$_3$ (0.25~eV/60~meV) is well within this range. 
(iii)The pseudogap disappears above 120~K, but the mIR peak 
is observed even at 690~K.     In the theoretical 
$\sigma(\omega)$, although the gap is fully developed only 
at $T \ll T_K$, the spectral weight below $\sim \Delta_{dir}$ 
begins to decrease at $\sim T_K$, as sketched in 
Fig.~4(d).\cite{mutou,jarrel,rozenberg}     
$T_K$ of YbAl$_3$ is $\sim$ 700~K,\cite{cornelius,PES} hence 
the observation of mIR peak up to 690~K is reasonable.
Degiorgi {\it et al.}\cite{degiorgi-mIR} have also argued 
that the mIR peaks observed for HF metals should 
correspond to $\Delta_{dir}$, and that their presence 
at $T \gg T^\ast$ is reasonable, based on considerations 
on the PAM somewhat different from those in the present work.

We have shown that the observation of both a pseudogap 
and a mIR peak with different characteristic temperatures 
can be qualitatively explained by our model, taking into 
account the indirect transitions.   However, some questions 
remain unanswered:    At first, 
the theories\cite{riseborough,logan} predict only a tail in 
$\sigma(\omega)$, as sketched in Fig.~4(c), decreasing from 
$\omega$=$\Delta_{dir}$ to vanish at $\Delta_{ind}$.    
Namely, {\it these theories cannot account for the 
presence of a shoulder}; its microscopic origin 
remains unclear at the present time.    
It may be necessary to consider effects not included 
in the simple PAM, such as the orbital degeneracy, the crystal 
field, and the underlying band structure.    
In the case of Kondo semiconductor, a realistic theory 
including these effects has recently been started,\cite{saso} 
to understand the gap formation more consistently.    
Secondly, the band picture sketched in Fig.~4(a) is 
well defined only at $T \leq T^\ast$.    At $T \sim T_K$, 
the quasiparticle lifetimes are too short to define such bands.   
The interpretation of mIR peak as ``optical absorptions across 
$\Delta_{dir}$'' is invalid at high $T$.    
Presently, there seems no such clear-cut description for the 
mIR peak at high $T$.    It probably results from a local 
charge excitation at each Yb site, but its microscopic 
character is unclear.     Note that a simple excitation 
from a Kondo singlet is unlikely, since such excitation should 
involve a change in the spin and should be negligibly weak 
in $\sigma(\omega)$.\cite{wooten}     
In addition, it is very remarkable that the inelastic 
neutron scattering experiment of YbAl$_3$ has revealed a 
pseudogap of about 30~meV at low temperatures.\cite{murani}    
The relation between the observed pseudogaps in the spin (neutron) 
and the charge (optical) excitation spectra is unclear.     
We expect that further advances in the theories of the 
$f$-electron systems should clarify these problems.

In conclusion, the $\sigma(\omega)$ spectra of the IV compound 
YbAl$_3$ have been measured over wide ranges of $\omega$ and $T$.   
They exhibited many characteristic features: an extremely 
narrow Drude peak, a pseudogap having a width of $\sim k_B T_K$, 
a pronounced mIR peak, and heavy-mass Fermi liquid 
properties.     These features and their $T$ dependences 
can be qualitatively understood based on PAM.   
In particular, the pseudogap and the mIR peak have been 
analyzed in terms of indirect and direct gaps within the PAM.   
To further establish the relation between the mIR peak/pseudogap 
and the $c$-$f$ hybridization, it will be useful to study 
$\sigma(\omega)$ as a function of tunable hybridization strength.    
For this purpose, IR experiments on IV/HF compounds 
under hydrostatic pressure are under way.

\section*{Acknowledgements}
We thank T. Watanabe and M. Matsunami for technical 
assistance, and T. Mutou for many useful comments 
on the theoretical aspects.    
We also acknowledge useful discussions with H. Harima, 
J. Lawrence, J.-M. Mignot, P. Riseborough, and T. Saso.    
This work is partly supported by a Grant-in-Aid from MEXT.   
T.E. is financially supported by Yukawa Foundation and 
Corning Japan.

\newpage

\begin{figure}
\begin{center}
\caption{(a) Optical reflectivity spectrum [$R(\omega)$] of 
YbAl$_3$ at 295~K as a function of photon energy 
($\omega$).  (b) $R(\omega)$ of YbAl$_3$ between 
8 and 690~K, and that of LuAl$_3$ at 295 K.   The spectra 
of YbAl$_3$ above 295~K and that of LuAl$_3$ are measured 
above 0.06~eV only.  }
\end{center}
\end{figure}

\begin{figure}
\begin{center}
\caption{(a)(b) Optical conductivity [$\sigma(\omega)$] 
spectra of YbAl$_3$.    In (b) the vertical broken line 
and the vertical arrow indicate the shoulder and the 
pseudogap, respectively, as discussed in the text.      
(c) $\sigma(\omega)$ at the pseudogap minimum at 27~meV 
(Pseudogap) and the intensity of the mIR peak 
integrated between 0.06 and 0.5~eV (mIR), normalized 
by those at 295~K, as a function of temperature ($T$).    
(d) A comparison between the low-energy $\sigma(\omega)$ 
and the dc conductivities, indicated at $\omega$=0 with 
circle (8~K), square (80~K), triangle (160~K), and cross 
(295~K).    The curves below 7~meV (the vertical broken 
line) are guide to the eye.   
}
\end{center}
\end{figure}

\begin{figure}
\begin{center}
\caption{
(a) Energy-dependent scattering rate $\tau^{-1}(\omega)$ 
of YbAl$_3$ as a function of the photon energy $\omega$. 
An $\tau^{-1}(\omega) \propto \omega^2$ dependence 
is shown for comparison.    
(b) Energy-dependent effective mass $m^\ast(\omega)$ 
relative to the bare optical mass $m_b$.   
}
\end{center}
\end{figure}

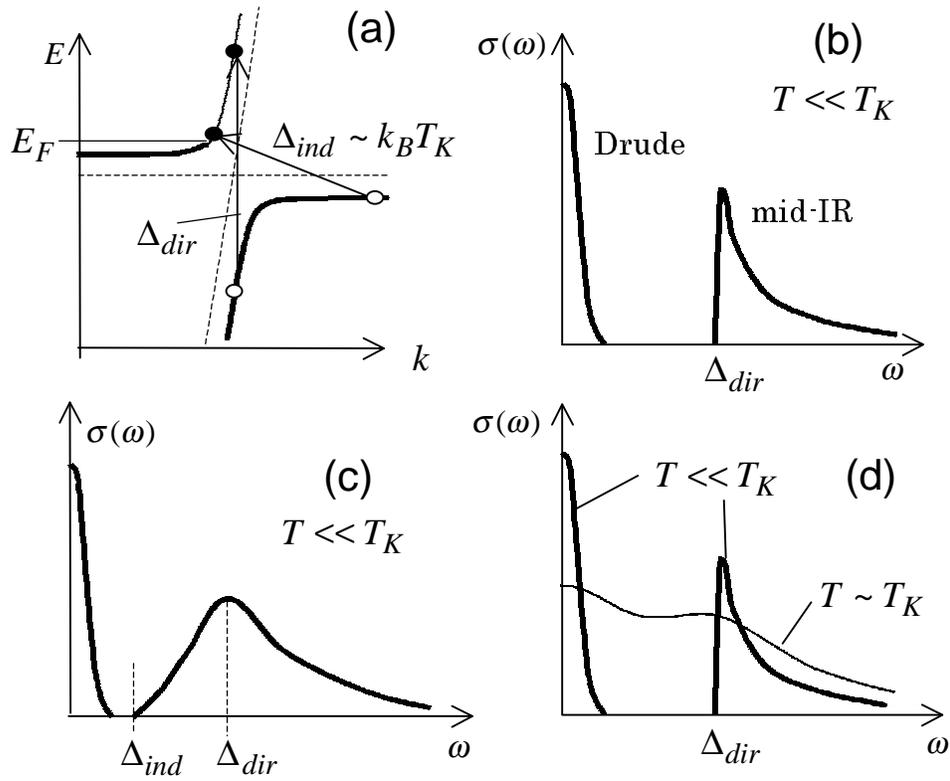
\begin{figure}
\begin{center}
\caption{
Illustration of the thoretical predictions by the PAM 
discussed in the text.  (a) $c$-$f$ hybridization states 
near the Fermi level ($E_F$).    The arrows indicate 
optical transitions across the direct ($\Delta_{dir}$) 
and the indirect ($\Delta_{ind}$) gaps.   
(b) $\sigma(\omega)$ at $T \ll T_K$ calculated with 
direct transitions only, and (c) that with both direct 
and indirect transitions.  
(d) Temperature dependence of $\sigma(\omega)$.   
Only direct transitions are included here.       
}
\end{center}
\end{figure}

\end{document}